\def\R{\item}

\documentclass[letterpaper, 11pt]{article} 
\usepackage[english]{babel} 
\usepackage[latin1]{inputenc}
\usepackage{amsmath,amsthm}
\usepackage{amssymb,latexsym}
\usepackage{graphics}
\usepackage{graphicx}

\newtheorem{prop}{Proposition}[section]
\newtheorem{defini}{Definition}[section]

\newcommand{\resumen}[2]{%
\begin{abstract}#1\\\\\noindent\textit{\textbf{Keywords and Phrases}: }#2\end{abstract}}

\title{\large\bf INFLATED BETA DISTRIBUTIONS\bigskip}
\author{
RAYDONAL OSPINA \\
Departamento de Estatística/IME-USP \\
Universidade de São Paulo. \\
Rua do Matão 1010, São Paulo/SP, 05508-090, Brazil. \\
{\tt e-mail:rospina@ime.usp.br }  \\
\\
SILVIA L.P. FERRARI \\ 
Departamento de Estatística/IME-USP \\
Universidade de São Paulo. \\
Rua do Matão 1010, São Paulo/SP, 05508-090, Brazil. \\
{\tt e-mail:sferrari@ime.usp.br} \\
 }
\date{}

\begin{document}
\maketitle
\centerline{\rule{14cm}{0.5pt}}
\resumen{
\noindent
This paper considers the issue of modeling fractional data observed on [0,1),
(0,1] or [0,1]. Mixed continuous-discrete distributions are proposed. The beta distribution is used
to describe the continuous component of the model since its density can have quite different
shapes depending on the values of the two parameters that index the distribution. Properties of the
proposed distributions are examined. Also, estimation based on maximum likelihood
and conditional moments is discussed. Finally, practical applications that employ real data
are presented.
}
{Beta distribution; inflated beta distribution; fractional data; maximum likelihood
estimation; conditional moments; mixture; proportions.}

\centerline{\rule{14cm}{0.5pt}}

\section{Introduction}
\label{intro}
\noindent Many studies in different areas involve data in the form of fractions, rates or
proportions that are measured continuously in the open interval $(0,1).$ However, frequently
the data contain zeros and/or ones.
In such cases, continuous distributions are not suitable for modeling the data.
In this work, we propose mixed continuous-discrete distributions to model data
that are observed on [0,1), (0,1] or [0,1]. The proposed distributions capture
the probability mass at 0, at 1 or both, depending on the case. For data observed
on $[0,1)$ or $(0,1]$ we use a mixture of a continuous distribution on $(0,1)$ and
a degenerate distribution that assigns non-negative probability to 0 or 1, depending
on the case. If the response variable is observed on the closed interval $[0,1] $ we use a
mixture of a continuous distribution on $(0,1)$ and the Bernoulli distribution, which
gives non-negative probabilities to 0 and 1. These models are special cases of the
class of {\it inflated models}. The word \emph{inflated} suggests that the probability
mass of some points exceeds what is allowed by the proposed model
(Tu,~2002). Some related works include Aitchison~(1955), Feuerverger~(1979),
Yoo~(2004), Heller, Stasinopoulos \& Rigby~(2006),  Cook, Kieschnick \& McCullough~(2004)
and Lesaffre, Rizopoulus \& Tsonaka~(2007).

The paper unfolds as follows. Section 2 presents the zero- and one-inflated beta
distributions and discusses some of their properties. Estimation based on maximum
likelihood and conditional moments is presented. Section 3 introduces the zero-and-one-inflated
beta distribution, some of its properties and estimation based on maximum likelihood and
conditional moments. In Section 4, Monte Carlo simulation studies are carried
out to examine the performance of the proposed estimators. Section 5 contains applications of the
proposed distributions and Tobit models to real data. For all the applications the
inflated beta distributions fitted the data better. Section 6 closes the paper with
concluding remarks.

\section{Zero- and one-inflated beta distributions}
\label{sec:1}

\noindent The beta distribution  is very flexible for modeling data that are measured in a
continuous scale on the open interval $(0,1)$ since its density has quite different shapes
depending on the values of the two parameters that index the distribution; see Johnson,
Kotz \& Balakrishnan~(1995, Chapter 25, Section 1), Kieschnick \& McCullough~(2003)
and Ferrari and Cribari--Neto~(2004).
The beta distribution with parameters $\mu$ and $\phi$ ($0 <\mu <1 $ and $\phi>0$),
denoted by ${\cal B}(\mu,\phi)$, has density function
\begin{equation}  
\label{beta}  
f(y;\mu,\phi)=\frac{\Gamma (\phi)}{\Gamma(\mu\phi)\Gamma((1-\mu)\phi)} \  
y^{\mu\phi-1}(1-y)^{(1-\mu)\phi-1},\quad y\in (0,1),  
\end{equation} where $\Gamma(\cdot)$ is the gamma function. If $y\sim{\cal B}(\mu,\phi)$,
then ${\rm E}(y)=\mu$ and 
${\rm Var}(y)={\rm V}(\mu)/(\phi+1),$ where ${\rm V}(\mu)={\mu(1-\mu)} $  denotes the
``variance function''. The parameter $\phi$ plays the role of a precision parameter
in the sense that, for fixed $\mu$, the larger the value of $\phi,$ the smaller the
variance of $y.$ Different values of the parameters generate different shapes
of the beta density (unimodal, `$U$', `$J$', inverted `$J$', uniform).

In practical applications the data may include zeros and/or ones. The beta distribution
is not suitable for modeling the data in these situations. If the data set contains zeros
or ones (but not both) its is natural to model the data using a mixture of two distributions:
a beta distribution and a degenerate distribution in a known value $c$, where $c=0$ 
or $c=1$, depending on the case. The cumulative distribution function  of the mixture
distribution is given by  
\begin{equation*}  
{\rm BI}_{c}(y;\alpha,\mu,\phi)=\alpha {\rm 1\!l}_{[c,1]}(y)+(1-\alpha)F(y;\mu,\phi),  
\end{equation*} 
where ${\rm 1\!l}_{A}(y)$ is an indicator function that equals 1 if $y\in A$ and
0 if $y\notin A$. Here, $F(\cdot;\mu,\phi)$ is the cumulative distribution function of
the beta distribution ${\cal B}(\mu,\phi)$ and $0 <\alpha <1$ is the mixture parameter.
The corresponding probability density function with respect to the measure generated by the
mixture\footnote{The probability measure $P$ corresponding to ${\rm BI}_{c}(y; \cdot),$
defined over the measurable space $((0,1)\cup\{c\},\mathfrak{B})$ where $\mathfrak{B}$ is
the class of all Borelian subsets of  $(0,1)\cup\{c\},$ is such that 
$P << \lambda+\delta_{c}$, with $\lambda$ representing the Lebesgue measure and
$\delta_{c}$ is a point mass  at c, i.e. $\delta_{c}(A)=1,$ if $c\in A$ and
$\delta_{c}(A)=0,$ if  $c\notin A,$ $A\in\mathfrak{B}.$ } is given by 
\begin{equation}  
\label{betamist}  
{\rm bi}_{c}(y;\alpha, \mu,\phi)=
\begin{cases}  
\alpha, & \text{if $y=c $}, \\  
(1-\alpha)f(y; \mu,\phi), & \text{if $y\in(0,1) $},  
\end{cases}  
\end{equation}
where  $f(y; \mu,\phi)$ is the  beta density \eqref{beta}. Note that $\alpha$ is
the probability mass at $c$ and represents the probability of observing 0 $(c=0)$
or 1 $(c=1)$.

\begin{defini}  
Let $y$ be a random variable that follows the inflated beta  distribution \eqref{betamist}.  
\begin{itemize}  
\item[1.]   
If $c=0,$ distribution \eqref{betamist} is called zero-inflated beta distribution
{\rm (BEZI)} and we write $y\sim {\rm BEZI}(\alpha,\mu,\phi).$
\item[2.]  
If $c=1,$ distribution \eqref{betamist} is called one-inflated beta distribution
{\rm (BEOI)} and we write $y\sim {\rm BEOI}(\alpha,\mu,\phi).$   
\end{itemize}  
\end{defini}
If $y\sim {\rm BEZI}(\alpha,\mu,\phi)$, then $\alpha=P(y=0)$ and if
$y\sim {\rm BEOI}(\alpha,\mu,\phi)$,
then $\alpha=P(y=1).$ Hence, those distributions allow us
to include a mass point at 0 or 1 in the beta distribution \eqref{beta}. 
 
The $r$th moment of $y$ and its variance can be written as
\begin{equation}  
\label{mediavar}
\begin{aligned}
{\rm E}(y^r)&=\alpha c+(1-\alpha)\mu_r, \ \ r=1,2,\ldots, \\
{\rm Var}(y)&=(1-\alpha)\frac{V(\mu)} {\phi+1}+\alpha(1-\alpha)(c-\mu)^2, \\  
\end{aligned}
\end{equation}
where $\mu_r=(\mu\phi)_{(r)}/(\phi)_{(r)}$, with $a_{(r)}=a(a+1)\cdots(a+r-1)$, is the $r$th
moment of the beta distribution \eqref{beta}. Note that ${\rm E}(y^r)$ is the weighted
average of the $r$th moment of the degenerate distribution
at $c$ and the corresponding moment of the beta distribution ${\cal B}(\mu,\phi)$
with weights $\alpha$ and $1-\alpha$, respectively. In particular,
${\rm E}(y)=\alpha c+(1-\alpha)\mu$.   
  
\begin{figure}[!htb]
\centering
\resizebox{0.9\textwidth}{!}{%
  \includegraphics[angle=270]{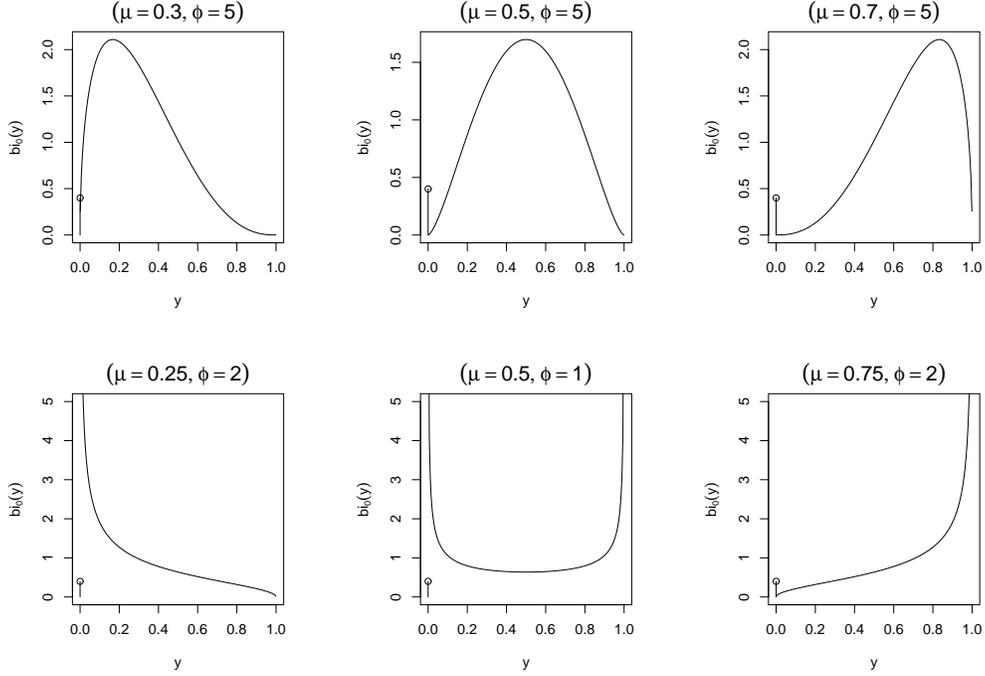}
}
\caption{BEZI densities for different values of $\mu$ and $\phi;$ $\alpha = 0.4$.}
\label{figbiz}       
\end{figure}

Figure \ref{figbiz} presents BEZI densities for different choices of $\mu$ and $\phi$ with
fixed $\alpha.$ Note that for all
$\mu$ and $\phi$ the BEZI distribution is asymmetrical because of the probability mass at 0.
Also, the BEZI density may be unimodal and may have `$J$',  `$U$', inverted `$J$' and uniform
shapes. In these graphs, the vertical bar with the circle above represents
$\alpha=P(y=0).$  Similarly, the BEOI distribution is asymmetrical because of the probability
mass at 1 and, for identical choices of the parameters, the BEZI and BEOI distributions have
the same functional shape on the interval $(0,1).$ However, they differ in
the mass point, being at 0 for the BEZI distribution and at 1 for the BEOI distribution.

\begin{prop}  
\label{teor1}  
The zero- and one-inflated beta distributions are three-parameter exponential family distributions
of full rank.
\end{prop}  
  
\begin{proof}
Let $\eta=(\eta_1,\eta_2,\eta_3),$ with $\eta_1=[\log(\alpha/(1-\alpha)) +B(\eta_2,\eta_3)], $
$\eta_2=\mu\phi$ and $\eta_3=(1-\mu)\phi,$ where $B(\eta_2,\eta_3)=
\log(\Gamma(\eta_2)\Gamma(\eta_3)/\Gamma(\eta_2+\eta_3)). $ Let $T(y)=(t_1(y),t_2(y),t_3(y)),$
where $t_1(y)={\rm 1\!l}_{\{c\}}(y),$ $t_2(y)= \log y$ if $y\in(0,1)$ and 0 if $y=c$ and
$t_3(y)= \log(1-y)$ if $y\in(0,1)$ and 0 if $y=c.$ Note that density \eqref{betamist} can
be written as  
\begin{equation}  
\label{famexpo}  
\exp\{\eta^\top T(y)-B^*(\eta) \} h(y),  
\end{equation} where $B^*(\eta)=\log\{1+\exp[\eta_1-B(\eta_2,\eta_3)] \} +B(\eta_2,\eta_3)$
is a real-valued function of $\eta$ and $h(y) = 1/\{y(1-y)\}$ if $y\in(0,1)$ and 1 otherwise
is a positive function defined over the set $(0,1)\cup\{c \}.$ The parameterization $\eta$
defines a one-to-one transformation which maps $\mathfrak{X}=\{(\alpha,\mu,\phi):(\alpha,\mu,\phi)
\in (0,1)\times (0,1)\times I \!\!R^+ \}$ onto $\mathfrak{D}=I \!\!R^+\times I\!\!R^+\times I\!\!R,$
i.e., the Jacobian of the transformation is nonzero for all $\eta \in \mathfrak{D},$ an open subset
of $I\!\!R^3.$ Additionally, neither the $t$'s nor the $\eta$'s satisfy linear constraints and the
parameter space contains a three-dimensional rectangle. Therefore, \eqref{famexpo} is the canonical
representation of the inflated beta distribution in the three-parameter exponential family of full
rank.
\end{proof}  
 
Let $y_1,\ldots,y_n$ be $n$ independent random variables, where each $y_t$ has density
\eqref{betamist}. A consequence of Proposition \ref{teor1} is that
$\sum_{t=1}^nT(y_t)=(T_1,T_2,T_3),$  with $T_1=\sum_{t=1}^n {\rm 1\!l}_{\{c\}}(y_t),$
$T_2=\sum_{t: y_t\in(0,1)} \log y_t$ and $T_3=\sum_{t: y_t\in(0,1)}\log (1-y_t),$ is a
complete (minimal) sufficient statistic (Lehmann \& Casella, 1998, Corollary 1.6.16
and Theorem 1.6.22).
 
The likelihood function for $\theta=(\alpha,\mu,\phi)$ given the sample $(y_1,\ldots,y_n)$ is
\begin{equation*}  
\label{verbetamis}  
L(\theta)=\prod_{t=1}^n{\rm bi}_{c}(y_t; \alpha,\mu,\phi) = L_1(\alpha)L_2(\mu,\phi),   
\end{equation*} where  
\begin{equation*}  
\begin{aligned}  
L_1(\alpha)&=\prod_{t=1}^n \alpha^{{\rm 1\!l}_{\{c\}}(y_t)}(1-\alpha)^{1-{\rm 1\!l}_{\{c\}}(y_t)}=
\alpha^{T_1}(1-\alpha)^{n-T_1}, \\  
L_2(\mu,\phi)&=\prod_{t=1}^nf(y_t;\mu,\phi)^{1-{\rm 1\!l}_{\{c\}}(y_t)}.
\end{aligned}  
\end{equation*}
The likelihood function $L(\theta)$ factorizes in two terms; the first term depends
only on $\alpha$ and the second, only on $(\mu,\phi)$. Hence, the parameters are separable
(Pace \& Salvan, 1997, p.\ 128) and maximum likelihood inference for $(\mu,\phi)$ can be
performed separately from that for $\alpha,$ as if the value of $\alpha$ were known, and vice-versa.

The log-likelihood function for the inflated beta distribution \eqref{betamist} is given by
$$\ell(\theta)=\log(L(\theta)) =
\ell_1(\alpha)+\ell_2(\mu,\phi), $$
where  
\begin{equation*}  
\begin{aligned}  
\ell_1(\alpha)&= T_1\log\alpha +(n-T_1)\log(1-\alpha), \\  
\ell_2(\mu,\phi)&= (n-T_1)\log\Bigg\{\frac{\Gamma(\phi)}{\Gamma(\mu\phi)\Gamma((1-\mu)\phi)}\Bigg\}+
T_2(\mu\phi-1)\\
&+T_3((1-\mu)\phi-1). \\  
\end{aligned}  
\end{equation*} 
The score function obtained by differentiating the log-likelihood function with respect to the
unknown parameters is $(U_{\alpha}(\alpha),U_{\mu}(\mu,\phi),U_{\phi}(\mu,\phi)),$ where
$U_{\alpha}(\alpha)={{T_1}}/ {{\alpha}} - {(n-T_1)} /(1 - \alpha),$
$U_{\mu}(\mu,\phi)=\phi\{(n-T_1)[\psi((1-\mu)\phi)-\psi(\mu\phi)] +T_2-T_3 \}$ and  
$U_{\phi}(\mu,\phi)=(n-T_1)[\psi(\phi)-\mu\psi(\mu\phi)-(1-\mu)\psi((1-\mu)\phi)]+
T_2\mu-T_3(1-\mu). $ The maximum likelihood (ML) estimator  of $\alpha$
is $\widehat\alpha=T_1/n$ and represents the proportion of zeros ($c=0$) or ones
($c=1$) in the sample. Since $\widehat\alpha$ is a function of a complete sufficient
statistic and is an unbiased estimator of $\alpha$, it is the uniformly minimum variance unbiased
estimator (UMVUE) of $\alpha$ (Lehmann \& Casella, 1998, Theorem 2.1.11); its variance
is given by ${\rm Var}(\widehat\alpha)=\alpha(1-\alpha)/n.$  The maximum likelihood estimators
of $\mu$ and $\phi$ are obtained from the equations $U_{\mu}(\mu,\phi)=0$ and
$U_{\phi}(\mu,\phi)=0,$ and do not have closed form. They can be obtained by
numerically maximizing the log-likelihood function $\ell_2(\mu,\phi)$ using a nonlinear
optimization algorithm, such as a Newton algorithm or a quasi-Newton algorithm; for details,
see Nocedal \& Wright (1999). Recently, the BEZI and BEOI distributions were incorporated
into the {\tt gamlss.dist} package in {\tt R} (Ospina, 2006). 

We can obtain estimators for $(\mu,\phi)$ based on conditional moments of $y$ given that
$y\in(0,1)$, which do not depend on $\alpha$. Observe that ${\rm E}(y \; | \; y\in(0,1))=\mu$ and
${\rm Var}(y \; | \; y\in(0,1))=\mu(1-\mu)/(\phi+1)$. For $T_1<n \ $\footnote{If $T_1=n$
(all observations equal $c$) the BEZI and
the BEOI distributions are not recommended.}, the solution of the system of equations
$(\overline{y},s^2)^\top=(\mu,\ {\mu(1-\mu)}/{(\phi+1)})^\top$, \ 
with $\overline{y}=\sum_{t: y_t\in(0,1)} y_t/ {(n-T_1)}$ and $s^2=\sum_{t: y_t\in(0,1)}
(y_t-\overline{y})^2/{(n-T_1)}$, gives the following closed-form estimators for $\mu$ and $\phi$:
$\widetilde\mu=\overline{y}$ and $\widetilde\phi = \{{\widetilde\mu(1-\widetilde\mu)}/s^2\}-1$.
Closed-form estimators of ${\rm E}(y^r)$ and ${\rm Var}(y)$ can be obtained by replacing 
$\alpha$, $\mu$ and $\phi$ by $\widehat \alpha$, $\widetilde \mu$ and $\widetilde \phi$
in (\ref{mediavar}).

The Fisher information matrix for the inflated beta distribution \eqref{betamist} is
\begin{equation}
\label{fisher1}
K(\theta)=
\begin{pmatrix}
\kappa_{\alpha\alpha} &0&0 \\
0&\kappa_{\mu\mu}&\kappa_{\mu\phi} \\
0&\kappa_{\phi\mu}&\kappa_{\phi\phi} \\
\end{pmatrix},
\end{equation} where  
$\kappa_{{\alpha\alpha}}=1/\{\alpha(1-\alpha )\}, $
$\kappa_{{\mu\mu}}      =(1-\alpha)\phi ^2 \{\psi'(\mu  \phi )+\psi'((1 -\mu)  \phi )\}, $
$\kappa_{{\mu\phi}}     = \kappa_{{\phi\mu}} = (1-\alpha)\phi\{\psi'(\mu \phi)\mu -
                       \psi'((1 -\mu)\phi)(1-\mu)\}$ and
$\kappa_{{\phi\phi}}    =(1-\alpha)\{\mu ^2 \psi'(\mu  \phi )+(1-\mu)^2\psi'((1 -\mu)  \phi ) -
                         \psi '(\phi )\}.$
Note that $K(\theta)$ does not depend on $c$. Also, $\alpha$ and $(\mu,\phi)$ are
globally orthogonal and  hence the corresponding components of the score vector are uncorrelated.
Since the inflated beta distribution \eqref{betamist} belongs to an exponential family
of full rank (see Proposition \eqref{teor1}), it follows that 
$\sqrt{n}(\widehat\theta - \theta) \ {\buildrel {\mathcal D} \over \rightarrow
\ {\mathcal N}_3}(0,K(\theta)^{-1})$,
with $K(\theta)$ given in \eqref{fisher1}
and that $\widehat\alpha $ and  $(\widehat \mu, \widehat\phi)$ are asymptotically independent.

If the interest lies in estimating a function of $\theta$, $r(\theta)$ say, 
the delta method (Lehmann \& Casella~1998,  \S \ 1.9) can be used to obtain the
asymptotic distribution of $r(\widehat\theta)$, the ML estimator of
$r(\theta)$. If $r(\theta)$ is differentiable, then
$\sqrt{n}(r(\widehat\theta) - r(\theta)) \ {\buildrel {\mathcal D} \over \rightarrow
\ {\mathcal N}}(0,\lambda(\theta))$, 
where $\lambda(\theta) = \dot{r}(\theta)^\top K(\theta)^{-1}\dot{r}(\theta)$
with $\dot{r}(\theta)=\partial r(\theta)/\partial \theta$. In particular, the 
maximum likelihood estimator of ${\rm E}(y)=\alpha c + (1-\alpha)\mu$ is
$\widehat\alpha c + (1-\widehat\alpha)\widehat\mu$ and  the variance of its normal limiting 
distribution is $(c-\mu)^2\kappa^{\alpha\alpha}+(1-\alpha)^2\kappa^{\mu\mu}$, where
$\kappa^{\alpha\alpha}=1/\kappa_{\alpha\alpha}$ and $\kappa^{\mu\mu}$ is
the $(2,2)$-element of $K(\theta)^{-1}$. In a similar fashion, ML estimation
of ${\rm Var}(y)$ can be performed.

\section{The zero-and-one-inflated beta distribution}
\label{sec:2}

The zero- and one-inflated distributions presented in Section 2 are not suitable for modeling
fractional data that contain both zeros and ones. For this situation, we propose a mixture
between a beta distribution  and a Bernoulli distribution.
Specifically, we assume that the cumulative distribution function of the random variable
$y$ is 
$$
{\rm BEINF}(y;\alpha,\gamma, \mu,\phi)=\alpha {\rm Ber}(y;\gamma)+(1-\alpha)F(y;\mu,\phi),
$$
where  ${\rm Ber}(\cdot;\gamma)$ represents the cumulative distribution function of a Bernoulli
random variable with parameter $\gamma $ and $F(\cdot;\mu,\phi)$ is the cumulative distribution
function of ${\cal B}(\mu,\phi).$ Here, $0 <\mu, \gamma,\alpha <1 $ and $\phi>0$,
$\alpha $ being the mixture parameter.

\begin{defini}  
Let $y$ be a random variable that assumes values in the closed interval $[0,1].$ We say that
$y $ has a zero-and-one-inflated beta distribution
{\rm (BEINF)} with parameters $\alpha$, $\gamma$, $\mu$ and $\phi$ if its
density function with respect to the measure generated by the
mixture\footnote{
The probability measure $P$ corresponding to ${\rm BEINF}(y; \cdot),$ defined over the measurable
space $([0,1],\mathfrak{B})$ where $\mathfrak{B}$ is the class of all Borelian subsets of $[0,1],$
is such that $P << \lambda+\delta_0+\delta_1,$ with $\lambda$ representing the Lebesgue measure and
$\delta_{c}$ is a point mass at $c,$ i.e. $\delta_{c}(A)=1,$ if $c\in A$ and 0, if $c\notin A,$
$A\in \mathfrak{B}.$} is given by
\begin{equation}
\label{betaIZU}
{\rm beinf}(y;\alpha,\gamma, \mu,\phi)=
\begin{cases}
\alpha(1-\gamma), &\text{\rm if } y=0 , \\
\alpha\gamma, &\text{\rm if } y=1 , \\
(1-\alpha)f(y; \mu,\phi), &\text{\rm if } y\in(0,1) ,
\end{cases}
\end{equation}
with $0<\alpha,\gamma,\mu<1$ and $\phi>0,$
where $f(y;\mu,\phi)$ the beta density function \eqref{beta}. We write
$y\sim{\rm BEINF}(\alpha,\gamma,\mu,\phi)$.  Note that, if $y\sim{\rm BEINF}(\alpha,\gamma,\mu,\phi)$, then
$P(y=0)=\alpha(1-\gamma)$ and $P(y=1)=\alpha\gamma$.
\end{defini}

After some algebra, the $r$th moment of $y$ and its variance can be written as   
\begin{equation}
\label{momentsbizu}
\begin{aligned}
{\rm E}(y^r)&=\alpha\gamma+(1-\alpha)\mu_r, \ \ r=1,2,\ldots,\\
{\rm Var}(y)&=\alpha V_1+(1-\alpha)V_2+\alpha(1-\alpha)(\gamma-\mu)^2, \\
\end{aligned}
\end{equation}
where $\mu_r $ is the $r$th moment of the beta distribution \eqref{beta},
$V_1=\gamma(1-\gamma)$ and 
$V_2={V}(\mu)/{(\phi+1)}.$
Note that ${\rm E}(y^r)$ is the weighted average of the $r$th moment of the Bernoulli
distribution with parameter $\gamma$ and the corresponding moment of the ${\cal B}(\mu,\phi)$
distribution with weights $\alpha$ and $1-\alpha$, respectively. 

\begin{prop}
\label{teor2}
The zero-and-one-inflated beta distribution given in \eqref{betaIZU} is a four-parameter exponential
family distribution of full rank.
\end{prop}                          
\begin{proof} Let $\eta=(\eta_1,\eta_2,\eta_3,\eta_4)$ with $\eta_1=[\log(\alpha/(1-\alpha)) -
M(\eta_2)+B(\eta_3,\eta_4)],$
$\eta_2=\break\hfill \log(\gamma/(1-\gamma)),$  
$\eta_3=\mu\phi$ and $\eta_4=(1-\mu)\phi$ where
$M(\eta_2)=\log(1+e^{\eta_2})$ and  
$B(\eta_3,\eta_4)=\log(\Gamma(\eta_3)\Gamma(\eta_4)/\Gamma(\eta_3+\eta_4))$  and let
$T(y)=(t_1(y),t_2(y), t_3(y),t_4(y))$ with
$t_1(y)={\rm 1\!l}_{\{0,1\}}(y),$ $t_2(y)=y{\rm 1\!l}_{\{0,1\}}(y),$ $t_3(y)=\log(y)$ if
$y\in(0,1)$ and 0 if $y\in\{0,1\}$
and $t_4(y)=\log(1-y)$ if $y\in(0,1)$ and 0 if $y\in\{0,1\}.$
Note that the BEINF density function \eqref{betaIZU} can be written as
\begin{equation}
\label{exp2}
\exp\{\eta^\top T(y)-B^*(\eta)\}h(y),
\end{equation}
where $B^*(\eta)=\log\{1+\exp[\eta_1+M(\eta_2)-B(\eta_3,\eta_4)] \} +B(\eta_3,\eta_4) $ is a
real-valued function of $\eta$ and $h(y)=1/\{y(1-y)\}$ if $y\in(0,1)$ and 1 if $y\in\{0,1\}.$
The parameterization $\eta$ defines a one-to-one transformation which maps
$\mathfrak{X}=\{(\alpha,\gamma, \mu,\phi):(\alpha,\gamma, \mu,\phi)\in
(0,1)\times(0,1)\times(0,1)\times I\!\!R^+\}$ onto $\mathfrak{D}=\{(\eta_1,\eta_2,\eta_3,\eta_4):
(\eta_1,\eta_2,\eta_3,\eta_4)\in I\!\!R\times I\!\!R\times I\!\!R^+ \times I\!\!R^+ \}.$
Additionally, neither the $t$'s nor the $\eta$'s satisfy linear constraints and the parameter
space contains a four-dimensional rectangle. Therefore, \eqref{exp2} is the canonical
representation of the BEINF distribution in the four-parameter exponential family of full rank.
\end{proof}  

Let $(y_1,\ldots,y_n)$ be a random sample of a BEINF distribution. Proposition \ref{teor2}
implies that $\sum_{t=1}^nT(y_t)=(T_1,T_2,T_3,T_4) $, with $T_1=\sum_{t=1}^n
{\rm 1\!l}_{\{0,1\}}(y_t),$ $T_2=\sum_{t=1}^n y_t {\rm 1\!l}_{\{0,1\}}(y_t),$
$T_3=\sum_{t: y_t\in(0,1)} \log(y_t)$ and $T_4=\sum_{t: y_t\in(0,1)} \log (1-y_t)$,
is a complete (minimal) sufficient statistic.
The likelihood function for $\theta=(\alpha,\gamma,\mu,\phi)$ given the sample
$(y_1,\ldots,y_n)$ is 
\begin{equation*}
\label{verobizu}
L(\theta)=\prod_{t=1}^n{\rm beinf}(y_t; \alpha, \gamma, \mu,\phi)=
L_1(\alpha)L_2(\gamma)L_3(\mu,\phi),
\end{equation*}
with  
\begin{equation*}
\begin{aligned}
L_1(\alpha) &= \quad \prod_{t=1}^n \alpha^{{\rm 1\!l}_{\{0,1\}}(y_t)}
(1-\alpha)^{1-{\rm 1\!l}_{\{0,1\}}(y_t)} = 
\alpha^{T_1}(1-\alpha)^{(n-T_1)},\\
L_2(\gamma)&= \prod_{t:y_t\in\{0,1\}}\gamma^{y_t}(1-\gamma)^{(1-y_t)}=
\gamma^{T_2}(1-\gamma)^{(T_1-T_2)},\\
L_3(\mu,\phi)&=\prod_{t:y_t\in(0,1)}f(y_t;\mu,\phi). 
\end{aligned}
\end{equation*} 
The likelihood function $L(\theta)$ factorizes in three terms, namely $L_1,$ $L_2$ and $L_3;$
$L_1$ depends only on $\alpha,$ $L_2,$ only on $\gamma$ and $L_3,$ only on $(\mu,\phi).$
Hence, $\alpha$, $\gamma$ and $(\mu,\phi)$ are separable parameters and maximum likelihood
inference for $\alpha$, $\gamma$ and $(\mu,\phi)$ can be performed independently. 
  
The log-likelihood function can be written as  
\begin{equation*}
\begin{aligned}
\ell(\theta)&=\log(L(\theta))=\ell_1(\alpha)+\ell_2(\gamma)+\ell_3(\mu,\phi),
\end{aligned}
\end{equation*}
where  
\begin{equation*}
\begin{aligned}
\ell_1(\alpha)&=
T_1\log\alpha +(n-T_ 1)\log(1-\alpha), \\
\ell_2(\gamma)&=
T_2\log\gamma +(T_1-T_2)\log(1-\gamma), \\
\ell_3(\mu,\phi)&=
(n-T_1)\log\Big\{\frac{\Gamma(\phi)}{\Gamma(\mu\phi)\Gamma((1-\mu)\phi)}\Big\}+T_3(\mu\phi-1)\\
&+T_4((1-\mu)\phi-1). \\
\end{aligned}
\end{equation*}
By differentiating  $\ell_1(\alpha)$ with respect to $\alpha,$ $\ell_2(\gamma)$ with respect to
$\gamma$ and $\ell_3(\mu,\phi)$ with respect to $\mu$ and $\phi$ we obtain the score vector 
$(U_{\alpha}(\alpha), U_{\gamma}(\gamma), U_{\mu}(\mu,\phi),  U_{\phi}(\mu,\phi))$, where   
$U_{\alpha}(\alpha)={{T_1}}/{{\alpha}} -{{(n-T_1)}}/{(1 - \alpha)},$ $U_{\gamma}(\gamma)=
{{T_2}}/{{\gamma}} -{{(T_1-T_2)}}/{(1 - \gamma)},$ 
$U_{\mu}(\mu,\phi)=\phi\{(n-T_1)[\psi((1-\mu)\phi)-\psi(\mu\phi)]+T_3-T_4\}$ and
$U_{\phi}(\mu,\phi)=(n-T_1)[\psi(\phi)-\mu\psi(\mu\phi)-(1-\mu)\psi((1-\mu)\phi)]+
\mu T_3-(1-\mu)T_4.$

It is easy to show that $\widehat\alpha ={T_1}/{n}$ and $\widehat\gamma ={T_2}/{T_1}$ ($0/0$
being regarded as 0) are the ML estimators of $\alpha$ and $\gamma$,
respectively. Here, $\widehat\alpha $ is the  proportion of zeros and ones in the sample
and $\widehat\gamma $ is the proportion of zeros among  the observations that equal zero
or one. Since $\widehat\alpha$ is a function of a complete sufficient statistic
and is an unbiased estimator of $\alpha$, it is the UMVUE of $\alpha$; its variance is given
by ${\rm Var}(\widehat\alpha)=\alpha(1-\alpha)/n.$ The ML estimators of
$\mu $ and $\phi$ are obtained as the solution of the nonlinear system of equations
$(U_{\mu}(\mu,\phi), U_{\phi}(\mu,\phi))=0 $. In practice, ML estimates
can be obtained through numerical maximization of the log-likelihood function $\ell_3(\mu,\phi)$
using a nonlinear optimization algorithm.
Closed-form estimators for $\mu$ and $\phi$, $\widetilde \mu$ and $\widetilde \phi$ say,
can be obtained using conditional moments of $y$ given
that $y\in (0,1)$, which do not depend neither on $\alpha$ nor on $\gamma$; see  Section 2.
Likewise, closed-form estimators of ${\rm E}(y^r)$ and ${\rm Var}(y)$ can be obtained by replacing 
$\alpha$, $\gamma$, $\mu$ and $\phi$ by $\widehat \alpha$, $\widehat\gamma$, $\widetilde \mu$
and $\widehat \phi$ in (\ref{momentsbizu}).

The Fisher information matrix for the parameters of the BEINF distribution can be written as 
\begin{equation}
\label{fisher2}
K(\theta)=
\begin{pmatrix}
\kappa_{\alpha\alpha} & 0&0&0 \\
0&\kappa_{\gamma\gamma}&0&0 \\
0&0&\kappa_{\mu\mu}&\kappa_{\mu\phi} \\
0&0&\kappa_{\phi\mu}&\kappa_{\phi\phi} \\
\end{pmatrix},
\end{equation}
where   $\kappa_{\alpha\alpha}=1/\{\alpha(1-\alpha)\},$ 
$\kappa_{\gamma\gamma}=\alpha/\{\gamma(1-\gamma)\},$
$\kappa_{{\mu\mu}}      =(1-\alpha)\phi ^2 \{\psi'(\mu  \phi )+\psi'((1 -\mu)  \phi )\}, $
$\kappa_{{\mu\phi}} = \kappa_{{\phi\mu}} = (1-\alpha)\phi\{\psi'(\mu \phi)\mu -\psi'((1 -\mu)\phi)
(1-\mu)\}$ and $\kappa_{{\phi\phi}}    =(1-\alpha)\{\mu ^2 \psi'(\mu  \phi )+(1-\mu)^2\psi'((1 -\mu)
\phi ) -\psi '(\phi )\}.$ Here, $\alpha,$ $\gamma$ and $(\mu,\phi)$ are orthogonal parameters
and, hence, the respective components of the score vector are uncorrelated.
Since the zero-and-one inflated beta distribution \eqref{betaIZU} belongs to an
exponential family of full rank (see Proposition \eqref{teor2}), it follows that 
$\sqrt{n}(\widehat\theta - \theta) \ {\buildrel {\mathcal D} \over \rightarrow
\ {\mathcal N}_4}(0,K(\theta)^{-1})$,
with $K(\theta)$ given in (\ref{fisher2}),
and $\widehat\alpha,$ $\widehat\gamma $ and $(\widehat \mu, \widehat\phi)$ are asymptotically
independent. 

The delta method (see Section 2) is useful for obtaining the asymptotic distribution of
the ML estimator of any differentiable function $r(\theta)$. For instance, if $r(\theta)=
{\rm E}(y)=\alpha\gamma+(1-\alpha)\mu$, the variance of the normal limiting distribution of
$\widehat{{\rm E}(y)}=r(\widehat\theta)$ is  
$\alpha^2{\kappa}^{{\gamma\gamma}}+(1-\alpha)^2{\kappa}^{{\mu\mu}}
+(\gamma-\mu)^2{\kappa}^{\alpha\alpha}$. Here, $\kappa^{\alpha\alpha}=1/\kappa_{\alpha\alpha},$
${\kappa}^{\gamma\gamma}=1/{\kappa}_{\gamma\gamma}$ and  ${\kappa}^{{\mu\mu}}$ is the
$(3,3)$-element of $K(\theta)$ given in (\ref{fisher2}).
  
There are other parameterizations of the BEINF distribution that can be useful.
For example, let $\gamma = \delta_1/\alpha $ and 
$\alpha=\delta_0+\delta_1.$ In this case, the BEINF density function can be written as  
\begin{equation}
\mathfrak{m}^{\star}(y; \delta_0, \delta_1,\mu,\phi)=
\begin{cases}
\delta_0, &\text{if $y=0$ }, \\
\delta_1, &\text{if $y=1$ }, \\
(1-\delta_0-\delta_1)f(y; \mu,\phi), &\text{if $y\in(0,1)$ },
\end{cases}
\end{equation}  
with $f(y; \mu,\phi)$ representing the beta density \eqref{beta}.
Here, the interpretation  of the parameters is more intuitive, since $\delta_0=P(y=0), $
$\delta_1=P(y=1)$ and $\mu, $ $\phi $ are the parameters of the  beta  distribution \eqref{beta}.   
However, this parameterization induces a restriction in the parameter space given by
$0 <\delta_0+\delta_1 <1.$  Fisher's information matrix for the BEINF distribution
in this parameterization
can be written as
\begin{equation*}
K(\theta)=
\begin{pmatrix}
\kappa_{\delta_0\delta_0} &\kappa_{\delta_0\delta_1}&0&0 \\
\kappa_{\delta_1\delta_0}&\kappa_{\delta_1\delta_1}&0&0 \\
0&0&\kappa_{\mu\mu}&\kappa_{\mu\phi} \\
0&0&\kappa_{\phi\mu}&\kappa_{\phi\phi} \\
\end{pmatrix},
\end{equation*}
where $\theta=(\delta_0,\delta_1,\mu,\phi)$,
$\kappa_{\delta_0\delta_0}={(1-\delta_1)}/{\delta_0(1-\delta_0-\delta_1)},$
$\kappa_{\delta_0\delta_1}=\kappa_{\delta_1\delta_0}=
{1}/{(1-\delta_0-\delta_1)},$
$\kappa_{\delta_1\delta_1}={(1-\delta_0)}/{\delta_1(1-\delta_0-\delta_1)},$
$\kappa_{{\mu\mu}} =(1-\alpha)\phi ^2 \{\psi'(\mu  \phi )+\psi'((1 -\mu)  \phi )\}, $
$\kappa_{{\mu\phi}} = \kappa_{{\phi\mu}} =
(1-\alpha)\phi\{\psi'(\mu \phi)\mu -\psi'((1 -\mu)\phi)(1-\mu)\}$ and
$\kappa_{{\phi\phi}}    =(1-\alpha)\{\mu ^2 \psi'(\mu  \phi )+(1-\mu)^2\psi'((1 -\mu)  \phi )
-\psi '(\phi )\}.$
Here $\kappa_{\delta_0\delta_1}\neq 0,$ thus indicating that $\delta_0$
and $\delta_1$ are not orthogonal parameters, and their respective \ components in the score
vector are correlated in \ contrast to the \ parameterization \ of the BEINF \ distribution given
in \eqref{betaIZU}. Recently, \ the BEINF distribution in this parameterization was
incorporated into the {\tt gamlss} package in {\tt R} (Stasinopoulos, Rigby \& Akantziliotou, 2006).

\section{Simulation results and discussion}
\label{sec:3}

We shall use Monte Carlo simulation to evaluate the finite sample performance of estimators
based on maximum likelihood (ML) and conditional moments (CM) for the BEZI and
BEINF distribution. For both distributions, the ML estimator of $\alpha$ is UMVUE. Hence, we do not show simulation results
for the estimation of such parameter. We focus our attention in estimation
of $\mu$, $\phi$, ${\rm E}(y)$ and ${\rm Var}(y)$; for the BEINF distribution,
the estimation of $\gamma$ is also considered. The parameters of the BEZI distribution
used in the numerical exercise were $\alpha=0.2,$ $\mu=0.1,$ and $\phi=2.$ 
For the BEINF distribution, $\alpha=0.2,$ $\gamma=0.3,$ $\mu=0.1$ and $\phi=2.$
The sample sizes considered were $n=10, 20, 50, 100, 500, 1000$ and
the number of Monte Carlo replications was $5,000.$
The ML estimates of $\mu$ and $\phi$
were obtained by maximizing the log-likelihood function using the BFGS method with analytical
derivatives; the BFGS quasi-Newton method is generally regarded as the best-performing nonlinear
optimization method (Mittelhammer, Judge and Miller, 2000, p.\ 199).
All simulations were performed using the {\tt Ox} matrix programming language (Doornik, 2006).  
  
Table \ref{tabs1} presents simulation results for the BEZI distribution. The estimated bias of the ML
estimators of  $\mu,$ ${{\rm E}(y)}$ and ${{\rm Var}(y)}$ are close to zero for all the sample sizes
considered. Also, the root mean square errors ($\sqrt{{\rm MSE}}$) of the ML and CM estimators of
$\mu,$ ${{\rm E}(y)}$ and ${{\rm Var}(y)}$ are similar. However, in small samples, the ML and CM
estimators of $\phi$ can be considerably biased, the CM estimator having much more pronounced bias
than the ML estimator. Additionally, the mean and the root mean square error of $\widetilde\phi$
is much larger than the corresponding figures obtained for $\widehat \phi$. For instance, for
$n=10$, the biases and the root mean square error are, respectively, 3.4 and 10.5 for the ML
estimator and 6.5 and 21.2 for the CM estimator. It is noteworthy that, for all the sample sizes,
the ML and CM estimators of $\phi$ have positive bias; however the variance of the response variable
is only slightly underestimated.

\begin{table}[!htb]
\begin{center}
{\scriptsize
\caption{Simulation results for the BEZI distribution; $\alpha=0.2,\mu=0.1,$ $\phi=2.0,$ ${\rm E}(y)=0.08$  and ${\rm Var}(y)=0.0256.$}
\label{tabs1}
}
\renewcommand{\tabcolsep}{0.45pc} 
\renewcommand{\arraystretch}{1.1} 
\begin{tabular}{@{}c|r|rr|rr|rr}\cline{3-8}
\multicolumn{1}{c}{} & \multicolumn{1}{c|}{} & \multicolumn{2}{|c|}{Mean} 
& \multicolumn{2}{|c|}{Bias} & \multicolumn{2}{|c}{$\sqrt{{\rm MSE}}$} \\ \cline{1-8}
Par.&$n$  & CM & ML & CM & ML & CM & ML  \\\hline

$           $&$10    $&$0.1009 $&$0.0945 $&$ 0.0009   $&$  -0.0055    $&$0.0637 $&$  0.0591  $\\
$           $&$20    $&$0.1005 $&$0.0971 $&$ 0.0005   $&$  -0.0029    $&$0.0432 $&$  0.0406  $\\
$\mu        $&$50    $&$0.1004 $&$0.0989 $&$ 0.0004   $&$  -0.0011    $&$0.0276 $&$  0.0262  $\\
$           $&$100   $&$0.0999 $&$0.0992 $&$-0.0001   $&$  -0.0008    $&$0.0194 $&$  0.0186  $\\
$           $&$500   $&$0.0998 $&$0.0996 $&$-0.0002   $&$  -0.0004    $&$0.0086 $&$  0.0083  $\\
$           $&$1000  $&$0.0997 $&$0.0997 $&$-0.0003   $&$  -0.0003    $&$0.0061 $&$  0.0058  $\\ \hline

$           $&$10    $&$8.5145 $&$5.3980$&$   6.5145 $&$  3.3980 $&$ 21.1880  $&$ 10.4990  $\\
$           $&$20    $&$3.6960 $&$2.9660$&$   1.6960 $&$  0.9660 $&$  4.6428  $&$  2.8064  $\\
$\phi       $&$50    $&$2.4009 $&$2.2750$&$   0.4009 $&$  0.2750 $&$  1.1773  $&$  0.8370  $\\
$           $&$100   $&$2.1840 $&$2.1371$&$   0.1840 $&$  0.1371 $&$  0.6793  $&$  0.5036  $\\
$           $&$500   $&$2.0340 $&$2.0292$&$   0.0340 $&$  0.0292 $&$  0.2487  $&$  0.1959  $\\
$           $&$1000  $&$2.0150 $&$2.0133$&$   0.0150 $&$  0.0133 $&$  0.1700  $&$  0.1342  $\\ \hline

$               $&$10    $&$0.0785 $&$ 0.0739$&$  -0.0015 $&$ -0.0061  $&$ 0.0503 $&$  0.0471  $\\
$               $&$20    $&$0.0802 $&$ 0.0775$&$   0.0002 $&$ -0.0025  $&$ 0.0354 $&$  0.0334  $\\
${{\rm E}(y)}   $&$50    $&$0.0804 $&$ 0.0792$&$   0.0004 $&$ -0.0008  $&$ 0.0228 $&$  0.0218  $\\
$               $&$100   $&$0.0799 $&$ 0.0794$&$  -0.0001 $&$ -0.0006  $&$ 0.0161 $&$  0.0155  $\\
$               $&$500   $&$0.0798 $&$ 0.0797$&$  -0.0002 $&$ -0.0003  $&$ 0.0072 $&$  0.0069  $\\
$               $&$1000  $&$0.0800 $&$ 0.0799$&$   0.0000 $&$ -0.0001  $&$ 0.0050 $&$  0.0048  $\\ \hline

$                $&$10    $&$0.0229 $&$ 0.0228$&$  -0.0027 $&$  -0.0028  $&$0.0229$&$   0.0203  $\\
$                $&$20    $&$0.0244 $&$ 0.0242$&$  -0.0012 $&$  -0.0014  $&$0.0167$&$   0.0153  $\\
${{\rm Var}(y)}  $&$50    $&$0.0253 $&$ 0.0252$&$  -0.0003 $&$  -0.0004  $&$0.0109$&$   0.0104  $\\
$                $&$100   $&$0.0254 $&$ 0.0253$&$  -0.0002 $&$  -0.0003  $&$0.0078$&$   0.0075  $\\
$                $&$500   $&$0.0255 $&$ 0.0255$&$  -0.0001 $&$  -0.0001  $&$0.0035$&$   0.0034  $\\
$                $&$1000  $&$0.0256 $&$ 0.0255$&$  -0.0000 $&$  -0.0001  $&$0.0024$&$   0.0024  $\\ \hline
\end{tabular}
\hfil
\end{center}
\end{table}

Table \ref{tabs2} summarizes simulation results for the BEINF distribution.
The ML estimator of $\gamma$ performs well if the sample size is not too small.
The  CM and ML estimators of $\mu$ are only slightly biased. On the other hand,
for very small samples (eg. $n=10$) the biases of the CM and ML estimators of
${{\rm E}(y)}$ and ${{\rm Var}(y)}$ are not negligible. We observed however, that
the ML estimator performs better than the CM estimator, both in terms of
bias and mean square error. Again, the CM and ML estimators of $\phi$ are
quite biased in small samples, the ML estimator performing better than the
CM estimator. For instance, for $n=20$ an $\phi=2$, the bias and the root mean square
error are approximately 1.0 and 2.4 for the ML estimator and 1.7 and 3.9 for the CM
estimator. 

In short, for the BEZI and the BEINF distributions the ML estimator of $\phi$ is more efficient
than the CM estimator, both estimators being quite biased for very small samples (eg. $n=10$).
On the other hand, the ML and CM estimators of the other parameters, and of ${{\rm E}(y)}$ and
${{\rm Var}(y)}$, have similar performances, both being almost unbiased if the sample is not
very small.

\begin{table}[!htb]
\begin{center}
{\scriptsize
\caption{Simulation results for the BEINF distribution; $\alpha=0.2,\gamma=0.3,\mu=0.1,$ $\phi=2.0,$ ${\rm E}(y)=0.14$  and ${\rm Var}(y)=0.0724.$}
\label{tabs2}
}
\renewcommand{\tabcolsep}{0.4pc} 
\renewcommand{\arraystretch}{1.1} 
\begin{tabular}{@{}c|r|rr|rr|rr}\cline{3-8}
\multicolumn{1}{c}{} & \multicolumn{1}{c|}{} & \multicolumn{2}{|c|}{Mean} 
& \multicolumn{2}{|c|}{Bias} & \multicolumn{2}{|c}{$\sqrt{{\rm MSE}}$} \\ \cline{1-8}
Par.&$n$  & CM & ML & CM & ML & CM & ML  \\\hline

$           $&$10    $&$        $&$0.4462$&$     $&$  0.1462  $&$  $&$ 0.1348 $\\
$           $&$20    $&$        $&$0.3882$&$     $&$  0.0882  $&$  $&$ 0.1616 $\\
$\gamma     $&$50    $&$        $&$0.3102$&$     $&$  0.0102  $&$  $&$ 0.1374 $\\
$           $&$100   $&$        $&$0.3015$&$     $&$  0.0015  $&$  $&$ 0.1047 $\\
$           $&$500   $&$        $&$0.2995$&$     $&$ -0.0005  $&$  $&$ 0.0466  $\\
$           $&$1000  $&$        $&$0.3011$&$     $&$  0.0011  $&$  $&$ 0.0323 $\\ \hline

$           $&$10    $&$0.1017  $&$ 0.0955 $&$  0.0017 $&$  -0.0045  $&$0.0661 $&$   0.0607  $\\
$           $&$20    $&$0.1003  $&$ 0.0969 $&$  0.0003 $&$  -0.0031  $&$0.0445 $&$   0.0418  $\\
$\mu        $&$50    $&$0.1006  $&$ 0.0992 $&$  0.0006 $&$  -0.0008  $&$0.0278 $&$   0.0265  $\\
$           $&$100   $&$0.0999  $&$ 0.0993 $&$ -0.0001 $&$  -0.0007  $&$0.0195 $&$   0.0186  $\\
$           $&$500   $&$0.1002  $&$ 0.1000 $&$  0.0002 $&$   0.0000  $&$0.0086 $&$   0.0083  $\\
$           $&$1000  $&$0.1001  $&$ 0.0999 $&$  0.0001 $&$  -0.0001  $&$0.0062 $&$   0.0060  $\\ \hline

$           $&$10    $&$10.9191 $&$   6.6869$&$  8.9191 $&$  4.6869$&$ 33.3939 $&$ 19.5225  $\\
$           $&$20    $&$ 3.7055 $&$   2.9909$&$   1.7055 $&$ 0.9909$&$  3.8645 $&$  2.3889 $\\
$\phi       $&$50    $&$ 2.4123 $&$   2.2821$&$   0.4123 $&$ 0.2821$&$  1.1997 $&$  0.8229 $\\
$           $&$100   $&$ 2.1824 $&$   2.1380$&$   0.1824 $&$ 0.1380$&$  0.6407 $&$  0.4891 $\\
$           $&$500   $&$ 2.0321 $&$   2.0210$&$   0.0321 $&$ 0.0210$&$  0.2424 $&$  0.1894 $\\
$           $&$1000  $&$ 2.0187 $&$   2.0115$&$   0.0187 $&$ 0.0115$&$  0.1726 $&$  0.1363 $\\ \hline

$               $&$10    $&$0.2002 $&$ 0.1956$&$ 0.0602 $&$ 0.0556  $&$ 0.0679 $&$  0.0655  $\\
$               $&$20    $&$0.1627 $&$ 0.1601$&$ 0.0227 $&$ 0.0201  $&$ 0.0524 $&$  0.0512  $\\
${{\rm E}(y)}   $&$50    $&$0.1423 $&$ 0.1411$&$ 0.0023 $&$ 0.0011  $&$ 0.0363 $&$  0.0357  $\\
$               $&$100   $&$0.1401 $&$ 0.1396$&$ 0.0001 $&$-0.0004  $&$ 0.0266 $&$  0.0263  $\\
$               $&$500   $&$0.1400 $&$ 0.1398$&$ 0.0000 $&$-0.0002  $&$ 0.0121 $&$  0.0119  $\\
$               $&$1000  $&$0.1402 $&$ 0.1401$&$ 0.0002 $&$ 0.0001  $&$ 0.0085 $&$  0.0084  $\\ \hline

$                $&$10    $&$0.1129 $&$ 0.1139 $&$   0.0405 $&$   0.0415  $&$0.0313 $&$   0.0306  $\\
$                $&$20    $&$0.0871 $&$ 0.0873 $&$   0.0147 $&$   0.0149  $&$0.0297 $&$   0.0293  $\\
${{\rm Var}(y)}  $&$50    $&$0.0727 $&$ 0.0727 $&$   0.0003 $&$   0.0003  $&$0.0232 $&$   0.0229  $\\
$                $&$100   $&$0.0717 $&$ 0.0717 $&$  -0.0007 $&$  -0.0006  $&$0.0177 $&$   0.0176  $\\
$                $&$500   $&$0.0721 $&$ 0.0721 $&$  -0.0003 $&$  -0.0003  $&$0.0080 $&$   0.0080  $\\
$                $&$1000  $&$0.0724 $&$ 0.0724 $&$   0.0000 $&$   0.0000  $&$0.0057 $&$   0.0056  $\\ \hline
\end{tabular}
\hfil
\end{center}
\end{table}

\section{Applications}
\label{sec:5}

This section contains three applications of inflated beta distributions to real
data. For the sake of comparison, we also fitted a Tobit model for
each data set. Computation for fitting inflated beta and Tobit models
was carried out using the packages {\tt gamlss} and {\tt VGAM} in the {\tt R}
software \ package (Ihaka \& Gentleman, 1996), \ respectively. \ In  \ 
{\tt gamlss}, we used the \ BEZI \ and \ the BEINF distributions implemented
by Ospina~(2006) and Stasinopoulos, Rigby \& Akantziliotou~(2006), respectively.

\begin{figure}[!htb]
\centering
\resizebox{1.03\textwidth}{!}{
  \includegraphics[angle=270]{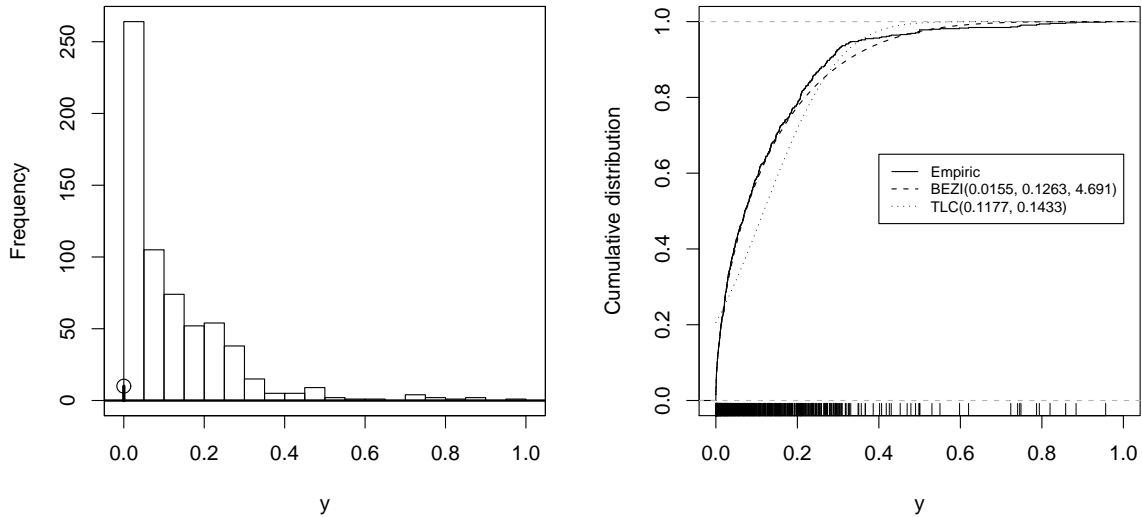}
}
\caption{Frequency histogram and estimated cumulative distribution
functions for the percentage of qualified nurses in Brazilian
municipal districts.}
\label{fig1}       
\end{figure}

The first application uses a data set of Brazilian indicators of qualified priority
services in 2000. The data were extracted from the {\it Atlas of Brazil Human Development}
database available at {\tt http://www.pnud.org.br/}. We modeled the percentage of qualified
nurses in 645 Brazilian municipal districts. The data set has zeros; some municipal districts
with high levels of poverty do not have qualified nurses.  The frequency histogram of the
data is presented in Figure \ref{fig1}. It has an inverted  `$J$' shape, a characteristic
easily modeled by the BEZI distribution. The vertical bar at zero represents the total
number of zeros in the sample.  We also considered a left censored Tobit model, by
assuming that $y_t=y^*_t$, if $y^*_t>0$, and  $y_t=0$, if $y^*_t\le 0$, where
$y^*_t \sim {\mathcal N}(\mu,\sigma^2)$ are independent random variables.
The ML estimates (standard errors in parentheses) for the parameters of the
BEZI distribution are $\widehat\alpha=0.0155 \ (0.0049),$ $\widehat\mu=0.1263 \ (0.0042),$ 
and $\widehat\phi=4.691 \ (0.220),$ and for the Tobit model,
$\widehat\mu=0.1177 \ (0.0060)$ and $\widehat\sigma=0.1433 \ (0.0040).$
The plot of the empirical distribution function of the data along with the estimated
cumulative distribution functions (see Figure 2) shows that only the BEZI distribution
fits the data well.

\begin{figure}[!htb]
\centering
\resizebox{1.03\textwidth}{!}{
  \includegraphics[angle=270]{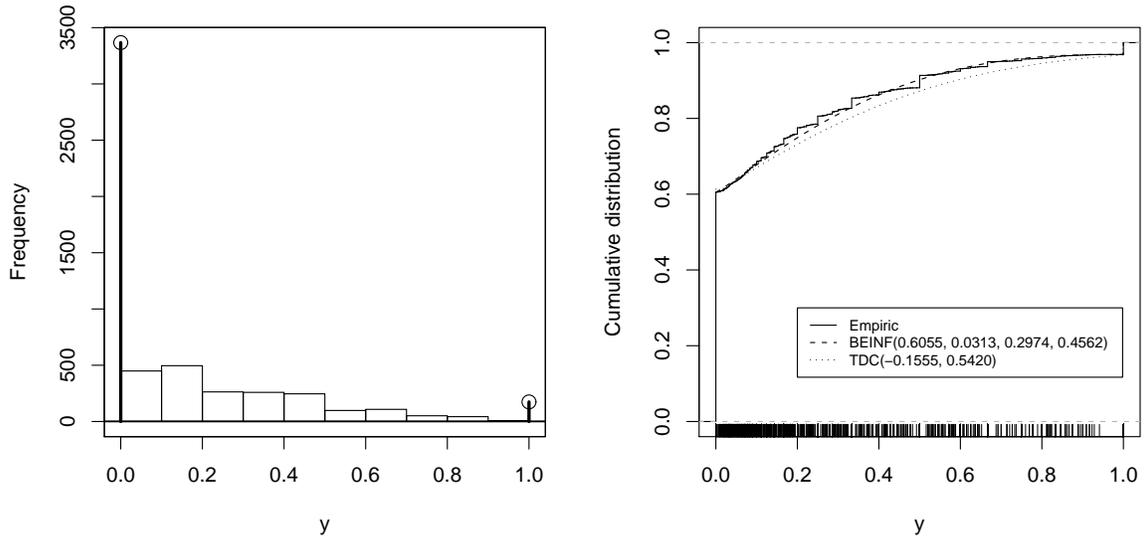}
}
\caption{Frequency histogram and estimated cumulative distribution functions for the
proportion of less than one year old infants that
died by unknown causes in Brazilian municipal districts in 2000.}
\label{fig3}       
\end{figure}

In the next application, we consider 5561 observations of proportions of less than one year
old infants that died by unknown causes in Brazilian municipal districts in 2000. The data
were obtained from the DATASUS database available at {\tt www.datasus.gov.br}. 
The data set contains 3364 zeros and 172 ones; the frequency histogram of the data is presented
in Figure \ref{fig3}.  For this data set we used a BEINF distribution under the parameterization
$(\delta_0,\delta_1,\mu,\phi).$ Also, we fitted a doubly censored Tobit model, i.e we assumed that
$y_t=y^*_t$, if $0<y^*_t<1$, \ $y_t = 0$, if $y^*_t\le 0$ and $y_t = 1$, if $y^*_t\ge 1$, where
$y^*_t \sim {\mathcal N}(\mu,\sigma^2)$ are independent random variables.
We obtained the following ML estimates: $\widehat\delta_0 = 0.6055 \ (0.0066),$
$\widehat\delta_1=0.0313 \ (0.0023), $ $\widehat\mu=0.2974 \ (0.0043)$ and
$\widehat\phi=0.4562 \ (0.0050)$ for the BEINF distribution, and
$\widehat\mu=-0.1555 \ (0.0088)$ and $\widehat\sigma=0.5420 \ (0.0085)$
for the Tobit model.
The empirical distribution and the BEINF and Tobit fitted cumulative distributions
are shown in Figure \ref{fig3}. By visual inspection, it becomes clear that only
the BEINF distribution is a suitable theoretical model to the data at hand.

\begin{figure}[!htb]
\centering
\resizebox{1.03\textwidth}{!}{%
  \includegraphics[angle=270]{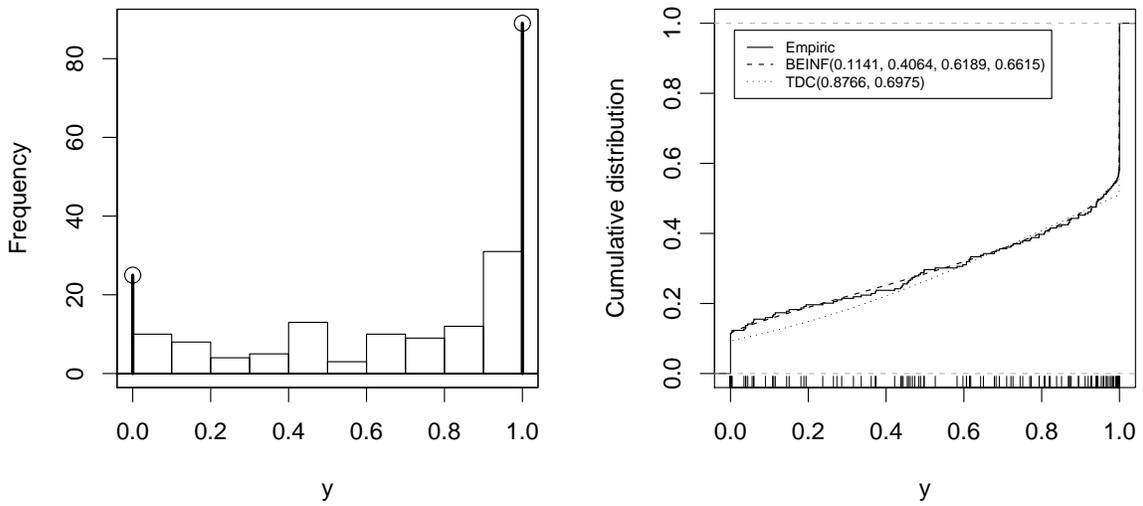}
}
\caption{Frequency histogram and estimated cumulative distribution functions;
coastal proximity data.}
\label{fig4}       
\end{figure}

Finally, we modeled the proportion of inhabitants who lived within a 200 km wide coastal strip
in 223 countries in the  year 2002. The data are supplied by the {\it Center for International
Earth Science Information Network } and are available at
{\tt http://sedac.ciesin.columbia.\break\hfill edu/plue/nagd/place.} Figure \ref{fig4} shows
that the histogram has a `$U$' shape. For these data we fitted a BEINF distribution under the parameterization
$(\delta_0,\delta_1,\mu,\phi)$ and a doubly censored Tobit model.
The ML estimates for the parameters of the BEINF distribution are
$\widehat\delta_0=0.1141\ (0.0215) ,$
$\widehat\delta_1=0.4064 \ (0.0332),$
$\widehat\mu=0.6189 \ (0.0279) $ and
$\widehat\phi=0.6615 \ (0.0204) .$ For the Tobit model, we obtained the following estimates:
$\widehat\mu=0.8766 \ (0.0518)$ and $\widehat\sigma=0.6975 \ (0.0368).$
Figure \ref{fig4} shows the empirical distribution and the estimated cumulative distribution curves.
Clearly, the Tobit model does not fit the data well. On the other hand, 
the empirical distribution and the BEINF estimated cumulative distribution curves are
quite close and we may conclude that the BEINF distribution is 
suitable to model the data. 

\section{Concluding remarks}
\label{sec:6}

The beta distribution is useful to model data that are measured
continuously on the open interval $(0,1)$. However, data sets that contain
zeros and/or ones cannot be modeled using a beta distribution.
A possible solution is to transform the response variable so that it assumes values
on the open unit interval. However, the use of transformations modifies the real
nature of the data and does not allow the direct interpretation of the parameters
in terms of the original response. 

In this paper, we propose mixed continuous-discrete distributions to
model data that are observed on $[0,1)$, $(0,1]$ or $[0,1]$.
The proposed distributions are ``inflated beta distributions'' in the sense
that the probability mass at 0 and/or 1 exceeds what is expected
by the beta distribution. Properties of the inflated beta distributions are
given. Also, estimation based on maximum likelihood and conditional moments is discussed
and compared using Monte Carlo simulation. Overall, we recommend maximum likelihood
estimation as the best choice.

An alternative to the inflated beta distributions is to assume that a latent
variable on $(0,1)$ gives rise to an observed response in $[0,1]$. This
approach has been suggested by Lesaffre, Rizopoulos and Tsonaka~(2007).
They assume that the $t$th observation is $y_t=r_t/N_t$, where
$r_t \sim {\rm Bin}(U_t,N_t)$ and the $U_t$'s follow a logit-normal distribution.
However, if the $N_t$'s are not known, as is the case of the examples
presented in Section 5, this model cannot be used. They also consider the 
situation where the response is assumed to be a coarsened version of a latent
variable with a logit-normal distribution on $(0,1)$. 
In other words, it is assumed that the logit transformed latent variable has a normal
distribution. The model we propose requires neither transformations nor the
inclusion of a latent variable. Tobit models, on the other hand, do not
require transformations but use a latent normal distributed variable.
The assumed normality of the latent variable does not allow the
Tobit models to be as flexible as the inflated beta distributions to model
fractional data. Additionally, the interpretation of the parameters
of Tobit models is rather difficult. For instance, the mean
of double censored Tobit responses involves the cumulative
distribution function and the probability density function of
a standard normal distribution; see Hoff~(2007, Section 4).

Three empirical applications using real data show that the inflated beta
distributions are quite flexible for modeling fractional data on
the closed or half-open unit interval. Also, for our data sets the Tobit
models did not work well \footnote{We are not claiming that the
inflated beta distributions always provide better fit than the Tobit models;
in this connection, see Hoff~(2007).}. 

We suggest that practitioners interested in modeling the behaviour 
of variables that assume values in the unit interval consider using a suitable
inflated  beta distribution whenever zeros and/or ones appear in the data set.

\section*{Acknowledgments}  
We gratefully acknowledge \ partial \ financial support \ from FAPESP and CNPq. We also thank
two anonymous referees for helpful comments and suggestions.
\section{References}
\label{referencias1}
\small

\begin{enumerate}
\footnotesize

\R{Aitchison, J.}~(1955). On the distribution of a positive random variable having a discrete
probability mass at the origin. {\it Journal of the American Statistical Association,} {\bf 50,} 901--908.

\R{Cook, D. O., Kieschnick, R. \& McCullough, B. D.}~(2004).
On the heterogeneity of corporate capital structures and its implications.
SSRN working paper; available at {\tt http://ssrn.com/abstract=671061}.

\R{Doornik, J.A.}~(2006). {\it Object-Oriented Matrix Language using {\tt Ox}}. 5n ed.  London: Timberlake Consultants Press.  

\R{Efron, B. \& Tibshirani, R. J.}~(1993). {\it An Introduction to the
Bootstrap}. New York: Chapman \& Hall.

\R{Feuerverger, A.}~(1979). On some methods of analysis for weather experiments. {\it Biometrika,} {\bf 66,} 665--668. 

\R{Ferrari, S.L.P. \& Cribari--Neto, F.}~(2004). Beta regression for modelling rates and proportions.
{\it Journal of Applied Statistics}, {\bf 31}, 799--815.

\R{Heller, G. Stasinopoulos, M. \& Rigby, B.}~(2006). The zero-adjusted inverse Gaussian distribution
as a model for insurance claims. {\it Proceedings of the 21th International Workshop on Statistical
Modelling}, J. Hinde, J. Einbeck, and J. Newell (Eds.) 226--233, Ireland, Galway.

\R{Hoff, A.}~(2007). Second stage DEA: Comparison of approaches for modelling the DEA score.
{\it European Journal of Operational Research,} {\bf 181}, 425--435.

\R{Ihaka, R. \& Gentleman, R.}~(1996).  R: A language for data analysis and graphics. 
{\it Journal of Computational and Graphical Statistics}, {\bf 5,} 299--314.

\R{Johnson, N., Kotz, S. \& Balakrishnan, N.}~(1995).
{\it Continuous Univariate Distributions}. 2nd  ed.
New York: John Wiley and Sons.

\R{Kieschnick R. \& McCullough, B. D.}~(2003). Regression analysis of variates
observed on (0,1): percentages, proportions, and fractions.
{\it Statistical Modelling,} {\bf 3}, 1--21.

\R{Lehmann, E. L. \& Casella, G.}~(1998).
{\it Theory of Point Estimation}, 2nd ed. New York: Springer.

\R{Lesaffre, E., Rizoupoulus, D. \& Tsonaka, S.}~(2007). The logistic-transform for bounded outcome
scores. {\it Biostatistics}, {\bf 8}, 72--85.

\R{Mittelhammer, R.C., Judge, G.G. \& Miller, D.J.}~2000. {\it Econometric Foundations}.
New York: Cambridge University Press.

\R{Nocedal, J. \& Wright, S. J.}~(1999). {\it Numerical Optimization.} New York: Springer.

\R{Ospina, R.}~(2006). The zero-inflated beta distribution for fitting a GAMLSS. Contribution to {\it R package gamlss.dist: Extra distributions to be used for GAMLSS modelling}. Available at {\tt http://r-project.org/\break\hfil CRAN/src/contrib/Descriptions/gamlss.dist.html}. 

\R{Pace L. \& Salvan, A.}~(1997). {\it Principles of Statistical Inference from a
Neo-Fisherian Perspective}. Singapore: World Scientific Publishing.

\R{Rigby, R. A. \&  Stasinopoulos D. M.}~(2005).
Generalized additive models for location, scale and shape (with discussion). 
{\it Applied Statistics}, {\bf 54}, 507--554.

\R{Stasinopoulos D. M., Rigby R. A. \& Akantziliotou C.}~(2006).
Instructions on how to use the {\tt GAMLSS} package in {\tt R}. 
Documentation in the current {\tt GAMLSS} help files
{\tt http://www.londonmet.ac.uk/gamlss/}.

\R{Tu, W.}~(2002). Zero inflated data. {\it Encyclopedia of Environmetrics,}   {\bf 4}, 2387--2391.

\R{Yoo, S.}~(2004). A note on an approximation of the mobile communications expenditures distribution
function using a mixture model. {\it Journal of Applied Statistics,} {\bf 31}, 747--752.

\end{enumerate}

\end{document}